\documentclass[10pt,twocolumn,twoside]{article}
\usepackage{amssymb}
\usepackage{amsmath}
\usepackage[dvips]{graphicx}
\usepackage[latin1]{inputenc}

 \hoffset-0.2in \voffset-0.0in
\begin{document}

\date{}
\title{Work and energy in inertial and non inertial reference frames}
\author{Rodolfo A. Diaz\thanks{%
radiazs@unal.edu.co}, William J. Herrera\thanks{%
jherreraw@unal.edu.co}, Diego A. Manjarrés\thanks{%
damanjarrnsg@unal.edu.co} \\
Departamento de Física. Universidad Nacional de Colombia. Bogotá, Colombia.}
\maketitle

\begin{abstract}
{\footnotesize It is usual in introductory courses of mechanics to develop
the work and energy formalism from Newton's laws. On the other hand,
literature analyzes the way in which forces transform under a change of
reference frame. Notwithstanding, no analogous study is done for the way in
which work and energy transform under those changes of reference frames. We
analyze the behavior of energy and work under such transformations and show
explicitly the expected invariance of the formalism under Galilean
transformations for one particle and a system of particles. The case of non
inertial systems is also analyzed and the fictitious works are
characterized. In particular, we show that the total fictitious work in the
center of mass system vanishes even if the center of mass defines a non
inertial frame. Finally, some subtleties that arise from the formalism are
illustrated by examples.}


\end{abstract}

\section{General formulation}

The behavior of the work and energy formulation under a change of reference
frame is not considered in most of the standard literature \cite{Klepp}, and
when considered, is limited to some specific cases \cite{AJPKap}. Recently,
the importance of Galilean invariance of the work energy theorem has been
highlighted by showing that the assumption of such an invariance implies the
impulse theorem \cite{EJPCamarca}. Thus we shall treat the problem in a
framework more general than previous approaches: the work energy theorem in
inertial and non inertial reference frames for an interacting non isolated
system of particles.\newline

Let us assume an inertial reference frame $\Sigma $ and another (inertial or
non inertial) reference frame $\Sigma ^{\prime }$. The origin of $\Sigma
^{\prime }$ moves arbitrarily with respect to $\Sigma $ but there is no
relative rotation between them. If $\mathbf{r}$ is the position of a
particle in $\Sigma $ then the position $\mathbf{r}^{\prime }\ $of the
particle in $\Sigma ^{\prime }$ reads

\begin{equation}
\mathbf{r}^{\prime }=\mathbf{r}-\mathbf{R,}  \label{rrp}
\end{equation}%
where $\mathbf{R}$ is the position of the origin of $\Sigma ^{\prime }$
measured by $\Sigma $. For the most general kinematics of such an origin
this relation becomes

\begin{equation}
\mathbf{r}^{\prime }=\mathbf{r}-\mathbf{R}_{0}-\int_{0}^{t}\mathbf{V(}t%
\mathbf{^{\prime })~}dt^{\prime },  \label{rrp2}
\end{equation}%
with $\mathbf{V}\left( t\right) $ denoting the velocity of $\Sigma ^{\prime
} $ with respect to $\Sigma $,

\begin{equation}
\mathbf{V(}t\mathbf{)=V}_{0}+\int_{0}^{t}\mathbf{A(}t\mathbf{^{\prime })~}%
dt^{\prime },  \label{vva}
\end{equation}%
where$\ \mathbf{R}_{0},\ \mathbf{V}_{0}$ are the initial values of $\mathbf{R%
}$, and $\mathbf{V}\ $respectively. $\mathbf{A}\left( t\right) $ refers to
the acceleration of $\Sigma ^{\prime }$ as observed by $\Sigma $. We work in
a scenario in which time intervals and masses are invariants. By
differentiation of Eq. (\ref{rrp}) we get

\begin{equation}
\mathbf{v}^{\prime }=\mathbf{v}-\mathbf{V(}t\mathbf{).}  \label{vvp}
\end{equation}

If we start with Newton's second law and use the fact that $\Sigma ^{\prime
} $ observes a fictitious force on the $j-$th particle of the form $\mathbf{F%
}_{fict}=-m_{j}\mathbf{A}\left( t\right) $, we get

\begin{equation}
m_{j}\frac{d\mathbf{v}_{j}^{\prime }}{dt}=\mathbf{F}_{j}-m_{j}\mathbf{A}(t),
\end{equation}%
$\mathbf{F}_{j}$ denotes the net \textbf{real} force on the $j-$particle
i.e. the sum of internal and external forces exerted on it. Taking the dot
product with $\mathbf{v}_{j}^{\prime }~dt\ $(on left) and $d\mathbf{r}%
_{j}^{\prime }\ $(on right), we obtain

\begin{equation}
m_{j}\mathbf{v}_{j}^{\prime }\cdot d\mathbf{v}_{j}^{\prime }=\left[ \mathbf{F%
}_{j}-m_{j}\mathbf{A}(t)\right] \cdot d\mathbf{r}_{j}^{\prime },
\end{equation}%
after summing over $j$, it leads to the work and energy theorem in $\Sigma
^{\prime }$

\begin{equation}
dK^{\prime }=dW^{\prime }.  \label{TFTENI}
\end{equation}

So the theorem of work and energy holds in the non inertial system $\Sigma
^{\prime }$ as long as the fictitious forces and their corresponding
fictitious works are included in $W^{\prime }$ as expected. Galilean
invariance is obtained by using $\mathbf{A}\left( t\right) =0$, from which
equality still holds and fictitious forces and works dissapear. It is useful
to obtain $dK^{\prime }$ and $dW^{\prime }$ in terms of variables measured
in the inertial frame $\Sigma $. We do this by using Eqs. (\ref{rrp2}, \ref%
{vva}, \ref{vvp}) and including the fictitious force on the $j-$th particle

\begin{eqnarray}
dK_{j}^{\prime } &=&m_{j}\mathbf{v}_{j}^{\prime }\cdot d\mathbf{v}%
_{j}^{\prime }  \notag \\
&=&m_{j}\left[ \mathbf{v}_{j}-\mathbf{V}\left( t\right) \right] \cdot \left[
d\mathbf{v}_{j}-\mathbf{A}\left( t\right) ~dt\right] ,  \label{dKjp1} \\
dW_{j}^{\prime } &=&\mathbf{F}_{j}^{\prime }\cdot d\mathbf{r}_{j}^{\prime } 
\notag \\
&=&\left[ \mathbf{F}_{j}-m_{j}\mathbf{A}\left( t\right) \right] \cdot \left[
d\mathbf{r}_{j}-\mathbf{V}\left( t\right) ~dt\right] .  \label{dWjp1}
\end{eqnarray}

Expanding these expressions we find

\begin{eqnarray}
dK_{j}^{\prime } &=&dK_{j}+dZ_{j},  \label{dKjp} \\
dW_{j}^{\prime } &=&dW_{j}+dZ_{j},  \label{dWjp}
\end{eqnarray}

\begin{equation}
dZ_{j}\equiv -\mathbf{V}\left( t\right) \cdot \mathbf{F}_{j}~dt-m_{j}\left[ d%
\mathbf{r}_{j}-\mathbf{V}\left( t\right) ~dt\right] \cdot \mathbf{A}\left(
t\right) .  \label{dZjp}
\end{equation}

From Eqs. (\ref{TFTENI}, \ref{dKjp}, \ref{dWjp}) and summing over $j\ $we
get $dK=dW$ as expected. In the galilean case $\mathbf{A}(t)=0$, we find

\begin{eqnarray}
dW_{j}^{\prime } &=&dW_{j}-\mathbf{V}\cdot \mathbf{F}_{j}~dt=dW_{j}-\mathbf{V%
}\cdot d\mathbf{P}_{j}~,  \notag \\
dW^{\prime } &=&dW-\mathbf{V}\cdot \mathbf{F}~dt=dW-\mathbf{V}\cdot d\mathbf{%
P,}  \label{dW dWp}
\end{eqnarray}%
where $d\mathbf{P}_{j}$ denotes the differential of linear momentum
associated with the $j-$particle. The second of these equations is obtained
just summing the first equation over $j$. It is easy to check that $d\mathbf{%
P}_{j}$ is the same for any inertial reference frame, but if $\Sigma
^{\prime }$ is non inertial we should take into account that $d\mathbf{P}%
_{j} $ is always measured by $\Sigma $ as can be seen in (\ref{dZjp}).

When $\Sigma ^{\prime }$ is non inertial, it is customary to separate $%
dW^{\prime }\ $in the work coming from real forces and the work coming from
fictitious forces. The work associated with fictitious forces can be easily
visualized in Eq. (\ref{dWjp1}) and we write

\begin{eqnarray*}
dW^{\prime } &=&dW_{real}^{\prime }+dW_{fict}, \\
dW_{fict} &=&-\mathbf{A}\left( t\right) \cdot \sum_{j=1}^{n}m_{j}\left[ d%
\mathbf{r}_{j}-\mathbf{V}\left( t\right) dt\right] ,
\end{eqnarray*}%
and Eq. (\ref{TFTENI}) can be rewritten in the following way

\begin{equation}
dW_{real}^{\prime }+dW_{fict}=dK^{\prime },  \label{TFTENI2}
\end{equation}%
which is similar in structure to the corresponding equation for forces. An
interesting case arises when $\Sigma ^{\prime }$ is attached to the center
of mass of the system. In that case, the total differential of fictitious
work reads

\begin{equation*}
dW_{fict}^{CM}=-\mathbf{A}_{C}\left( t\right) \cdot \sum_{j}m_{j}d\mathbf{r}%
_{j}^{\prime },
\end{equation*}%
with $\mathbf{A}_{C}$ denoting the acceleration of the center of mass. Since
the masses $m_{j}\ $are constant and using the total mass $M$ we get%
\begin{equation}
dW_{fict}^{CM}=-M\mathbf{A}_{C}\left( t\right) \cdot d\left( \frac{%
\sum_{j}m_{j}\mathbf{r}_{j}^{\prime }}{M}\right) =0.  \label{WfictCM}
\end{equation}%
The term in parenthesis vanishes because it represents the position of the
center of mass around the center of mass itself. So \textbf{the total
fictitious work in the center of mass system vanishes even if such a system
is non inertial}. Notice however that the total fictitious force does not
necessarily vanish, neither the fictitious work done over a specific
particle.

It is worth emphasizing that the reference frame attached to the center of
mass posseses some particular properties: (a)\ the relation between angular
momentum and torque $d\mathbf{L}_{CM}/dt=\mathbf{\tau }_{CM}$ holds even if
the CM system is non inertial, (b) the fictitious forces do not contribute
to the total external torque \cite{RMF}, (c) the fictitious forces do not
contribute to the total work as proved above. We point out however, that
this situation is different when the CM frame rotates with respect to an
inertial frame \cite{RMF}.

It is clear that the case of one particle arises by supressing the $j$ index
and only external forces appear on the particle. The case in which $\Sigma
^{\prime }$ is inertial (galilean transformation) appears when $\mathbf{A}=0$%
, and in that case we see that (a) the fictitious forces and works
dissappear, (b) the fundamental theorem holds in the same way as appears in $%
\Sigma $, that is $dW_{real}^{\prime }=dK^{\prime }$. It is worth remarking
that when $\Sigma ^{\prime }$\ is attached to the center of mass the
relation $dW_{real}^{\prime }=dK^{\prime }$ also holds even if $\Sigma
^{\prime }$ is non inertial.

\section{Examples}

There are other subtleties on this formulation that could be enlightened by
some examples

\textbf{Example 1}: Consider a block pushed across a frictionless table by a
constant force\textbf{\ }$\mathbf{F}$ through a distance $L$ starting at
rest (as observed by $\Sigma $). So $\Delta K=W$ implies$\ mv^{2}/2=FL$ and
the time taken is $t=mv/F$. But the table is in the dining car of a train
travelling with constant speed $V$ (measured by$\ \Sigma ^{\prime }$)\ in
the direction of $\mathbf{F}$. From the point of view of $\Sigma ^{\prime }$%
, the work done on the block and its change in the kinetic energy read%
\begin{eqnarray*}
W^{\prime } &=&F\left( L+Vt\right) =F\left( L+mvV/F\right) =FL+mvV \\
\Delta K^{\prime } &=&\frac{1}{2}m\left( V+v\right) ^{2}-\frac{1}{2}mV^{2}=%
\frac{1}{2}mv^{2}+mvV
\end{eqnarray*}%
which shows explicitly that $\Delta K=W$ implies $\Delta K^{\prime
}=W^{\prime }$. The reader can check that we find the same result by using
Eq. (\ref{dW dWp}).

\begin{figure}[tbh]
\begin{center}
\includegraphics[width=5.5cm]{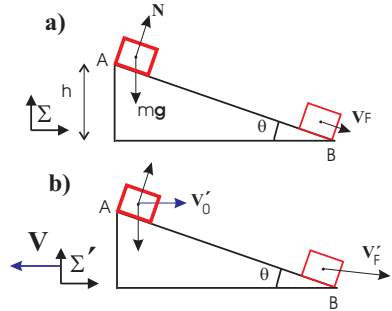}
\end{center}
\caption{\emph{Illustration of example 2. Force diagrams of a block over a
wedge, (a) with respect to }$\Sigma ,\ $\emph{(b) with respect to }$\Sigma
^{\prime }$\emph{. }$\mathbf{N,}$ \emph{m}$\mathbf{g}$ \emph{\ denote the
normal and gravitational forces respectively.}}
\label{fig:wedge}
\end{figure}

\textbf{Example 2}:\ Consider a block of mass $m$ sliding over a
frictionless wedge. In the $\Sigma $ frame, the block is at height $h$ and
at rest when $t=0$ (see Fig. \ref{fig:wedge}). The block will be our
physical system of interest. Assume that the reference frame $\Sigma
^{\prime }$ is another inertial system traveling with velocity $\mathbf{V}=-u%
\mathbf{i}$ toward left (see Fig. \ref{fig:wedge}). We intend to calculate
the work on the block and its final speed in the $\Sigma ^{\prime }$ frame.
For this, we apply Eq. (\ref{dW dWp}) in a finite form%
\begin{eqnarray*}
W^{\prime } &=&W-\mathbf{V}\cdot \mathbf{I}=mgh-\left( -u\mathbf{i}\right)
\cdot m\left[ \mathbf{v}_{f}-\mathbf{v}_{0}\right] \\
&=&mgh+mu\mathbf{i}\cdot \mathbf{v}_{f}~,
\end{eqnarray*}%
where $\mathbf{I}$ denotes impulse. By arguments of energy in $\Sigma $ we
see that $v_{F}=\sqrt{2gh}$ so that%
\begin{equation}
W^{\prime }=mgh+mu\sqrt{2gh}\cos \theta .  \label{trabajocorrecto0}
\end{equation}%
It can be checked by calculating explicitly the work due to each force, that
the extra term in $W^{\prime }$ owes to the fact that \textbf{the normal
force contributes to the work} in $\Sigma ^{\prime }$. It is because the
trajectory of the block in $\Sigma ^{\prime }$ is not perpendicular to the
normal force. Since the theorem of work and energy is valid in $\Sigma
^{\prime }$ and all forces are conservative, we can use the conservation of
mechanical energy applied to points $A$ and $B$ to get%
\begin{equation}
\frac{1}{2}mu^{2}+mgh+mu\sqrt{2gh}\cos \theta =\frac{1}{2}mv_{F}^{\prime 2},
\end{equation}%
and solving for $v_{F}^{\prime }$ we find%
\begin{equation}
v_{F}^{\prime }=\left( u^{2}+2gh+2u\sqrt{2gh}\cos \theta \right) ^{1/2}.
\label{velocidadprima}
\end{equation}%
It is easy to check the consistency of Eq. (\ref{velocidadprima}) since $%
\mathbf{v}_{0}^{\prime }=u\mathbf{i}$ and $\mathbf{v}_{F}^{\prime }\ $can be
obtained by taking into account that $\mathbf{v}_{F}^{\prime }=\mathbf{v}%
_{F}+u\mathbf{i}$ and using $v_{F}=\sqrt{2gh}$ from which we obtain Eq. (\ref%
{velocidadprima}).\ One of the main features illustrated by this problem is
that the work done by a force depends on the reference frame, and
consequently its associated potential energy (if any). In the case in which
both systems are inertial, forces are the same in $\Sigma $ and $\Sigma
^{\prime }$. However, works done by each force can change because the
trajectories are different in each system. For this particular problem, from
the point of view of $\Sigma ^{\prime }\ $the weight works in the same way
as observed by $\Sigma $, but the normal force does work which is opposite
to the observations in $\Sigma $. What really matters is that the work and
energy theorem holds in both\footnote{%
If we applied naively the conservation of energy by assigning the
traditional potential energies for $m\mathbf{g}$ and $\mathbf{N}$ ($mgh$ and
zero respectively), we would obtain the mechanical energies $%
E_{A}=mgh+\left( 1/2\right) mu^{2}$ and $E_{B}=\left( 1/2\right)
mv_{F}^{\prime 2}$. Using conservation of mechanical energy and applying Eq.
(\ref{velocidadprima}) we obtain that such an equality can only holds for
the particular cases $u=0$ or $\theta =0$.\ Such a contradiction comes from
an incorrect use of the potential energies when changing the reference
frame. Recalling that the potential energy associated with a constant force $%
\mathbf{F\ }$(in $\Sigma $)\ is $-\mathbf{F\cdot r}+const$, and taking into
account that $\mathbf{N}$ and $m\mathbf{g}$ are constant (in both frames),
then suitable potential energies for both forces in $\Sigma ^{\prime }\ $can
be constructed by using potential energies of the form $-\mathbf{F\cdot r}%
^{\prime }+const$.}. Finally, it is worth saying that although the
constraint force in this example produces a real work on the block in $%
\Sigma ^{\prime }$, it is a consequence of the motion in time of the wedge
as observed by $\Sigma ^{\prime }$, therefore it does not produce a virtual
work in the sense of D'Alembert principle \cite{Goldstein}, hence such a
principle holds in $\Sigma ^{\prime }$ as well.

\begin{figure}[tbh]
\begin{center}
\includegraphics[width=4.5cm]{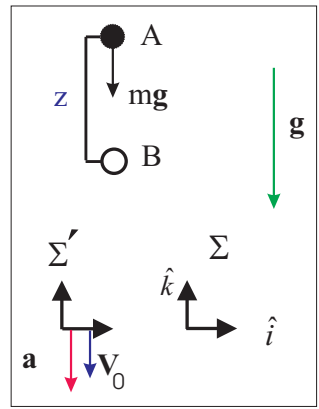}
\end{center}
\caption{\emph{Illustration of example 3.\ A ball in free fall with respect
to }$\Sigma $. $\Sigma ^{\prime }$\emph{\ has a uniform acceleration with
respect to }$\Sigma $.}
\label{fig:freefall}
\end{figure}


\textbf{Example 3}: Consider a ball that with respect to $\Sigma \ $is in
vertical free fall in a uniform gravitational field $\mathbf{g}$ starting at
rest. Assume that $\Sigma ^{\prime }$ is non inertial with acceleration $%
\mathbf{A}=-a\mathbf{k}$ and initial velocity $\mathbf{V}_{0}=-V_{0}\mathbf{k%
}$ with respect to $\Sigma $ (see Fig. \ref{fig:freefall}). The total work$\ 
$on the ball from $A\ $to\ $B$ measured by $\Sigma ^{\prime }$ involves real
and fictitious forces and can be obtained by integrating Eq. (\ref{dWjp1})%
\begin{eqnarray*}
W^{\prime } &=&m\left( a-g\right) ~\mathbf{k\cdot }\left\{ \int_{\mathbf{r}%
_{A}}^{\mathbf{r}_{B}}d\mathbf{r+}\int_{0}^{T}\left( V_{0}+at\right) ~%
\mathbf{k}~dt\right\} , \\
W^{\prime } &=&m\left( g-a\right) \left( z-V_{0}T-aT^{2}/2\right) ,
\end{eqnarray*}%
where $z$ is the distance covered by the ball from $A$ to $B$ as observed by 
$\Sigma $, and $T$ the time in covering such a distance. When $g=a$ then $%
W^{\prime }=0$. This fact can be seen from the equivalence principle, since
in the case in which $\mathbf{a}=\mathbf{g}$ the gravitational equivalent
field associated with the non inertial system cancels out the external
uniform field, so the force measured by such a system (and so the work)
vanishes. Finally, the reader can check that the fundamental theorem of work
and energy holds in $\Sigma ^{\prime }\ $for this case, as long as we use
the work $W^{\prime }$ which includes the fictitious work.

\section{Conclusions}

We have examined the behavior of the work and energy formulation for a
system of particles under a change of reference frame. We show explicitly
the galilean covariance of the work and energy theorem and show the way in
which such a theorem behaves in a non inertial frame. It is worth pointing
out that the form of the theorem is preserved when going to a non inertial
traslational frame as long as the fictitious works are included. In
addition, we found that when the reference frame is attached to the center
of mass, the total fictitious work is always null such that the work and
energy theorem is held without the inclusion of fictitious works even if
such a frame is non inertial. Finally, we illustrate the fact that after a
change of reference frame, the work done for each forces also changes (even
if the transformation is galilean). In consequence, the corresponding
potential energies should be changed when they exist. In particular, we show
that normal forces could produce work in some inertial reference frames.

\section*{ACKNOWLEDGMENTS}

We acknowledge the useful suggestions of two anonymous referees. We also
thank División de Investigación de Bogotá (DIB) from Universidad Nacional de
Colombia, for its financial support.

\appendix

\section{Suggestions for readers}

\begin{figure}[tbh]
\begin{center}
\includegraphics[width=7.0cm]{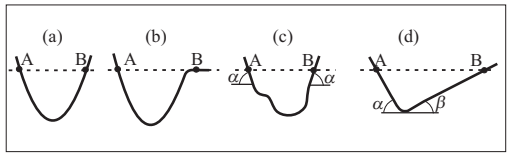}
\end{center}
\caption{\emph{Some configurations of a ball sliding over a frictionless
surface. The ball travels from point\ }$A\ $\emph{to point\ }$B$,$\ $\emph{%
and such points are fixed with respect to the surface.}}
\label{fig:ejemplos}
\end{figure}

For the reader to assimilate the principal features exposed, we propose the
following situations and questions,\newline

\begin{itemize}
\item Let us consider two balls of masses $m_{1}$, $m_{2}$ in free fall in a
uniform gravitational field with respect to a frame attached to the ground $%
\Sigma $. The mass $m_{1}$ starts at rest from a height $h$ while the mass $%
m_{2}$ is thrown upward from the ground on the same vertical with an initial
velocity $\mathbf{v}_{0}$. We ask the reader to show explicitly that the
total work measured by the center of mass system $\Sigma _{C}$ is null and
therefore the total kinetic energy remains constant. The fact that the
fictitious work is zero is a general property for $\Sigma _{C}$ as stated in
Eq. (\ref{WfictCM}), while the fact that the work done for real forces
vanishes is a particular characteristic of this problem.

\item As it has been noted previously, when we change of reference frame, an
amount of work associated with forces that do not do work in the initial
frame can appear. In the galilean case this amount of work is easily
evaluated from the net impulse. Consider a block of mass $m$ sliding over
different arrangements as depicted in Fig. 3. For all cases, the block is at
rest in the position $A$ with respect to the $\Sigma $ frame, and travels
from $A$ to $B$. Another inertial frame $\Sigma ^{\prime }$ is introduced,
which moves with speed $V$ to the left. Without making any explicit
calculations. For which cases does the normal force perform net work between
the points $A\ $and$\ B\ $in $\Sigma ^{\prime }$?.

\item For our second example, sketch adequate potential energies for the
block in both $\Sigma $ and $\Sigma ^{\prime }$.\ Does the potential energy
exhibit the same behaviour in both frames?, Why or Why not?. Further,
calculate explicitly the net work on the block observed by $\Sigma ^{\prime
} $. This is another way of checking that Eq. (\ref{trabajocorrecto0}) holds.

\item Imagine the very typical problem of a block sliding on a frictionless
surface, starting at a height $h$ and ending at $h=0$. We inmediately put $%
mgh=mv^{2}/2$. However, the conservation of mechanical energy requires to
use the potential energy associated with the net force on the block, while
in this case we are using the potential energy associated with an applied
force which clearly differs from the net force, under what circumstances is
the typical result correct?
\end{itemize}

\end{document}